\documentclass[11pt,aps,nofootinbib,prd,aps,epsf,floats,axodraw,amsmath,amssymb,amsfonts]{revtex4}
\usepackage{amsmath, amssymb}
\newcommand{\mathsym}[1]{{}}

\usepackage{graphicx}% Include figure files
\usepackage{amsmath}
\usepackage{amssymb}
\usepackage{bm}% bold math
\newcommand{\be}{\begin{equation}}
\newcommand{\ee}{\end{equation}}
\newcommand{\bea}{\begin{eqnarray}}
\newcommand{\eea}{\end{eqnarray}}

\newcommand{\rem}[1]{}
\newsavebox{\PSLASH}
 \sbox{\PSLASH}{$p$\hspace{-1.8mm}/}
 
\renewcommand{\theequation}{\thesection.\arabic{equation}}
\newcounter{saveeqn}
\newcommand{\add}{\addtocounter{equation}{1}}
\newcommand{\alpheqn}{\setcounter{saveeqn}{\value{equation}}%
\setcounter{equation}{0}%
\renewcommand{\theequation}{\mbox{\thesection.\arabic{saveeqn}{\alph{equation}}}}}
\newcommand{\reseteqn}{\setcounter{equation}{\value{saveeqn}}%
\renewcommand{\theequation}{\thesection.\arabic{equation}}}

 \newsavebox{\notrightarrow}
 \sbox{\notrightarrow}{$\to$\hspace{-4mm}/}
 
 \newsavebox{\PARTIALSLASH}
 \sbox{\PARTIALSLASH}{$\partial$\hspace{-1.6mm}/}
 
 \newsavebox{\ASLASH}
 \sbox{\ASLASH}{$A$\hspace{-2.1mm}/}
 
 \newsavebox{\KSLASH}
 \sbox{\KSLASH}{$k$\hspace{-1.8mm}/}
 
 \newsavebox{\LSLASH}
 \sbox{\LSLASH}{$\ell$\hspace{-1.8mm}/}
 
 \newsavebox{\QSLASH}
 \sbox{\QSLASH}{$q$\hspace{-1.8mm}/}
 
 \newsavebox{\DSLASH}
 \sbox{\DSLASH}{$D$\hspace{-2.2mm}/}
 
 \newsavebox{\DbfSLASH}
 \sbox{\DbfSLASH}{${\mathbf D}$\hspace{-2.8mm}/}
 
 \newsavebox{\DELVECRIGHT}
 \sbox{\DELVECRIGHT}{$\stackrel{\rightarrow}{\partial}$}
 
 \newcommand{\blue}{\IfColor{\textCadetBlue}{}}
\newcommand{\black}{\IfColor{\textBlack}{}}
\newcommand{\red}{\IfColor{\textRed}{}}
\newcommand{\green}{\IfColor{\textOliveGreen}{}}
\newcommand{\lila}{\IfColor{\textRedViolet}{}}

\begin{document}
\begin{flushright}
 [math-ph]
\end{flushright}
\title{Gauge Fixing Invariance\\ and\\ Anti-BRST Symmetry
}

\author{Amir Abbass Varshovi}\email{ab.varshovi@sci.ui.ac.ir/amirabbassv@ipm.ir/amirabbassv@gmail.com}

\affiliation{
   Department of Mathematics, University of Isfahan, Isfahan, IRAN.\\
   School of Physics, Institute for Research in Fundamental Sciences (IPM).\\
                                     Tehran-IRAN}
\begin{abstract}
       \textbf{Abstract\textbf{:}} It is shown that anti-BRST invariance in quantum gauge theories can be considered as the quantized version of the symmetry of  classical gauge theories with respect  to different gauge fixing mechanisms.
\end{abstract}
 \maketitle
%%%%%%%%%%%%%%%%%%%%%%%%
\section{Introduction}\label{introduction}
\indent  Since the advent of BRST quantization, it has been shown that beside the BRST symmetry the quantum Lagrangian is subjected to a further symmetry, known as anti-BRST invariance, through a significant role of anti-ghost field. This emergent symmetry was induced by anti-BRST transformation \cite{Curci, Ojima} which was similar to that of BRST but in particular the ghost field is replaced with the anti-ghost in variation of gauge and matter fields. Though, this symmetry turned out to be a useful tool for constraining possible terms in the action and in simplifying relations between Green’s functions, it has however been far from clear if there is any case in which it is indispensable, and thus its meaning has remained so far somewhat mysterious \cite{Henneaux1, Henneaux2, Weinberg, Bertlmann, Bonora, Binosi, Varshovi}.\\

\par In fact, the most mysterious subject of this symmetry is due to its geometric background. Particularly, although Yang-Mills theories and all their elaborations including ghost field and BRST symmetry are well understood by means of differential geometry the anti-BRST invariance and the anti-ghost field are still suffering of an ambiguity in their geometric origins. Really, although the geometry of anti-BRST was studied in various methods including some sophisticated algebra-geometric tools such as supermanifold formalisms \cite{Henneaux1, Henneaux2, Henneaux3, Henneaux4, Henneaux5, Hull, Barnich, Upadhyay}, gerbes \cite{Bonora, Bonora2} or axial extensions of principal bundles \cite{Varshovi}, but despite of the BRST structure there is still no consensus on the geometric origion of this Faddeev-Popov's accidentally emergent symmetry.\\

\par In fact, this lack of consensus is an inevitable consequence of having no idea for the classical background of anti-BRST symmetry \cite{Weinberg, Bonora}. More precisely, it seems at least through with some viewpoints that the anti-BRST invariance is an extra symmetry beside the BRST one for just one classical symmetry of gauge transformations \cite{Alvarez-Gaume}. In particular, by possessing the classical version of a quantum feature one can build up its geometric model by transcending the classical settings to those of the geometric structures of infinite dimensional functional spaces of quantum fields, the procedure which is applicable more or less in most geometric methods of mathematical quantum physics \cite{Bertlmann, Nash}. Therefore, the indispensable property for an infinite dimensional quantum structure in the theory of quantum fields to be compatibly modeled with geometric tools is to have a finite dimensional geometric counterpart in classical physics or quantum mechanics, the case of which has no realization for anti-BRST symmetry and anti-ghost.\\

\par Since generally there is considered no extra significant symmetry in classical gauge theories beside the gauge invariance, it seems that anti-BRST structure needs more sophisticated settings to be appeared in formulations rather than those of the geometry of finite dimensional manifolds\footnote{such as the principal bundles of classical gauge theories.}. That is anti-BRST invariance needs an infinite dimensional setting to be modeled thoroughly even in its classical version. Actually, anti-BRST symmetry is in its essence an infinite dimensional structure which ceases to be appeared within the geometry of classical gauge theories. But, by the way, we naturally expect this symmetry to have a trace on classical gauge theories despite of its exotic nature \cite{Weinberg, Alvarez-Gaume}.\\

\par In this article we will claim and show that the invariance of a gauge theory with respect to different gauge fixing procedures is the classical version of quantum anti-BRST symmetry. It is obviousely an infinite dimensional symmetry even in classical model of Yang-Mills theories. The aspects of this symmetry appears in quantum Lagrangian only after the gauge is fixed in Faddeev-Popov's approach just similar to BRST invariance. More precisely, as we will show by details, although incorporating gauge fixing term in quantum Lagrangian may lead to restriction of vastly huge space of gauge fields to a fairly smaller space of roots of gauge fixing function $f$ \footnote{This is the case in sharp or strict gauge fixing methods such as the Landau's gauge.}, it provides a set of new transformations to quantum fields due to continuous variation of $f$. This set of new transformations is geometrically shown to be that of anti-BRST. However, since the classical theory is physically consistent, we expect after quantization the theory be invariant under anti-BRST transformations. \\

\par To explain the main idea we should basically emphasize that the invariance of observables in Yang-Mills theories with respect to different gauge fixing methods provides an independent classical symmetry in such gauge theories beside the well known invariance of gauge transformations. Thus, there may be two probably relevant questions;\\
\begin{itemize}
\item Where the classical symmetry of gauge fixing invariance is encoded in quantized Yang-Mills theories?
\item Why there are two quantum symmetries, the BRST and the anti-BRST, versus just one classical gauge invariance?
\end{itemize}

\par These two paradoxical questions leads us naturally to the main idea of this article;\\
\begin{eqnarray}
\begin{array}{*{20}{c}}
  \emph{\emph{Is the classical symmetry of gauge fixing invariance emerged}}  \\
   \emph{\emph{as}} \\
   \emph{\emph{anti-BRST symmetry of quantized Yang-Mills gauge theories?}}\\
\end{array}
\end{eqnarray}
\\

\par In the next section we will derive the anti-BRST transformation by considering the gauge fixing invariance as its classical version. In the third section we will give a simple proof for validity of our model and then discuss and explain the topological necessity of anti-BRST symmetry beside the BRST one for all quantized Yang-Mills theories.\\

\par We should emphasize here that in this article we try to give a geometric model for anti-BRST symmetry of quantized Yang-Mills theorie via the Faddev-Popov quantization approach. Therefore, although there may be various similarities amoung all gauge theories such as the Yang-Mills, $p$-form gauge theories, the gravity, that of strings, ... , most of our elaborations are specified to Yang-Mills gauge theories. Hence, all through in this article, by gauge theories we mean those of Yang-Mills unless otherwise is stated clearly.

%%%%%%%%%%%%%%%%%%%%%%%%%%%%%%%%%%%%%
%
%%%%%%%%%%%%%%%%%%%%%%%%%%%%%%%%%%%%%%%

%%%%%%%%%%%%%%%%%%%%%%%%%%%%%%%%%%%%%%%%

\par
\section{Gauge Fixing and Anti-BRST}
\setcounter{equation}{0}
\par Let $(P,\pi,X)$ be a principal $G$-bundle for a Lie group $G$ with Lie algebra $\frak{g}$. Here, we may consider $X=P/G$ as the space-time manifold $\mathbb{R}^4$ and also $G$ to be compact. Thus, for simplicity $P$ could be assumed to be topologically a trivial fibration. Now set $\widetilde{P}$ as the space of all Cartan connection forms over $P$. The set of all smooth functions $\widetilde{g}:X\to G$ acts on $\widetilde{P}$ (from right) with;
\begin{eqnarray} \label {1} 
\widetilde{p}.\widetilde{g}=\widetilde{g}^{-1} \emph{\emph{d}}\widetilde{g}+Ad_{\widetilde{g}^{-1}} (\widetilde{p})  ,
\end{eqnarray}

\noindent for $\widetilde{p} \in \widetilde{P}$, and for $\emph{\emph{d}}$ the exterior derivative operator on $P$, where on the right hand side of (\ref{1}) $\widetilde{g} \circ \pi$ is replaced with $\widetilde{g}$ for simplicity. Let’s define $\widetilde{G}$ to be the collection of all fixed point preserving functions $\widetilde{g}:X\to G$. In fact, according to equation (\ref{1}), any elelment of $\widetilde{G}$ is conventionally refered to as a gauge transformation. We note that there is a family of natural epic group homomorphisms $\phi_x:\widetilde{G}\to G$, with $\phi_x(\widetilde{g})=\widetilde{g}(x)$, for any $x\in X$ except for the point of which we specialized for fixed point preserving gauge transformations\footnote{Symbolically we consider this point as the infinity point of $S^4$ due to compactification of $\mathbb{R}^4$.} \cite{Singer}. In the same way, the Lie algebra of $\widetilde{G}$ as a Lie group consists of smooth functions $\widetilde{\alpha}:X\to \frak{g}$, known as infinitesimal gauge transformations. We show this Lie algebra with $\widetilde{\frak{g}}$. Similarly, Lie group homomorphisms $\phi_x$ lead to Lie algebra homomorphsms in a natural way; $\emph{\emph{d}}\phi_x:\widetilde{\frak{g}}\to \frak{g}$, with $\emph{\emph{d}}\phi_x(\widetilde{\alpha})=\widetilde{\alpha}(x)$. Despite of definition of $\widetilde{G}$, it doesn't contain $G$ as a subgroup provided fixed point preserving property is applied to elements of $\widetilde{G}$, since due to this property any $\widetilde{g}\in \widetilde{G}$ has to represent the identity elelment of $G$ at the specified fixed point.\\

\par It is well-known that $\widetilde{G}$ acts freely on $\widetilde{P}$, leading to a principal $\widetilde{G}$-bundle; $(\widetilde{P},\widetilde{\pi},\widetilde{X})$ for $\widetilde{X}=\widetilde{P}/\widetilde{G}$. This principal bundle may not have a trivial topology even if $(P,\pi,X)$ is trivial \cite{Singer}, the fact of which is well understood for non-Abelian gauge group $G$ as Gribov ambiguity in the literature \cite{Gribov}. In fact, for any non-Abelian group $G$ the Gribov ambiguity will appear as a topological feature \cite{Singer}.\\

\par Now, let $\widetilde{\widetilde{P}}$ be the set of all Cartan connection forms over $\widetilde{P}$. Similarly, any smooth function $\widetilde{\widetilde{g}}:\widetilde{X}\to \widetilde{G}$ acts on $\widetilde{\widetilde{P}}$ in the same way as (\ref {1}), i.e;
\begin{eqnarray} \label {2} 
\widetilde{\widetilde{p}}.\widetilde{\widetilde{g}}=\widetilde{\widetilde{g}}^{-1} \tilde{\emph{\emph{d}}} \widetilde{\widetilde{g}}+Ad_{\widetilde{\widetilde{g}}^{-1}} (\widetilde{\widetilde{p}})  ,
\end{eqnarray}

\noindent for $\widetilde{\widetilde{p}} \in \widetilde{\widetilde{P}}$, and for $\tilde{\emph{\emph{d}}}$ the exterior derivative operator on $\widetilde{P}$, while again we use $\widetilde{\widetilde{g}}$ instead of $\widetilde{\widetilde{g}} \circ \widetilde{\pi}$. We show the group of all functions $\widetilde{\widetilde{g}}$ with $\widetilde{\widetilde{G}}$ \footnote{Here we avoide of unnecessary restriction to fixed point preserving condition. However, one can similarly consider a principal $\widetilde{\widetilde{G}}$-bundle $(\widetilde{\widetilde{P}},\widetilde{\widetilde{\pi}},\widetilde{\widetilde{X}})$ for some suitable restriction on elements of $\widetilde{\widetilde{G}}$ in order to make the action (\ref{2}) to be free. Actually, here we do not need such rigorous structures to make a hierarchy of principal bundles. This topic will be discussed and appeared in \cite{Varshovi-next2}.}. The elelments of $\widetilde{\widetilde{G}}$ are referred to as gauge fixing transformations for the reason which will be cleared in the following. Accordingly, we call $\widetilde{\widetilde{G}}$ the gauge fixing transformation group.\\

\par Similarly, there is a family of epic homomorphisms $\phi_{\widetilde{x}}:\widetilde{\widetilde{G}}\to \widetilde{G}$, with $\phi_{\widetilde{x}}(\widetilde{\widetilde{g}})=\widetilde{\widetilde{g}}(\widetilde{x})$, for any $\widetilde{x}\in \widetilde{X}$. The Lie algebra of $\widetilde{\widetilde{G}}$ as a Lie group consists of smooth functions $\widetilde{\widetilde{\alpha}}:\widetilde{X}\to \widetilde{\frak{g}}$. We show this Lie algebra with $\widetilde{\widetilde{\frak{g}}}$. Let's refer to the elements of $\widetilde{\widetilde{\frak{g}}}$ as infinitesimal gauge fixing transformations. Accordingly, Lie group homomorphisms $\phi_{\widetilde{x}}$ lead to Lie algebra homomorphsms in a natural way; $\emph{\emph{d}}\phi_{\widetilde{x}}:\widetilde{\widetilde{\frak{g}}}\to \widetilde{\frak{g}}$, with $\emph{\emph{d}}\phi_{\widetilde{x}}(\widetilde{\widetilde{\alpha}})=\widetilde{\widetilde{\alpha}}(\widetilde{x})$. In fact, if we do not impose any extra structures on $\widetilde{\widetilde{G}}$, such as fixed point preserving condition, $\widetilde{\widetilde{G}}$ will contain $\widetilde{G}$ as a proper Lie subgroup.\\

\par According to introduced structures define;
\begin{eqnarray} \label {3}
\begin{array}{*{20}{c}}
   i:\widetilde{P}\times \widetilde{G}\times \widetilde{\widetilde{G}}\to \widetilde{P}~~~~~~~~~~~~~~~  \\
   ~~~~~~~~~~~~~~(\widetilde{p},\widetilde{g},\widetilde{\widetilde{g}})~~~\mapsto \widetilde{p}.\widetilde{g}.\phi_{\widetilde{x}}(\widetilde{\widetilde{g}})=\widetilde{p}.\widetilde{g}\widetilde{\widetilde{g}}(\widetilde{x}),  \\
\end{array}
\end{eqnarray}

\noindent for $\widetilde{x}=\widetilde{\pi}(\widetilde{p})\in \widetilde{X}$. In fact, we read from (\ref{3});
\begin{eqnarray} \label {4}
i(\widetilde{p},\widetilde{g},\widetilde{\widetilde{g}})=\widetilde{\widetilde{g}} (\widetilde{x})^{-1}\emph{\emph{d}}\widetilde{\widetilde{g}}(\widetilde{x})+Ad_{\widetilde{\widetilde{g}} (\widetilde{x})^{-1}}(\widetilde{g}^{-1} \emph{\emph{d}}\widetilde{g})+Ad_{\widetilde{\widetilde{g}} (\widetilde{x})^{-1} \widetilde{g}^{-1}} (\widetilde{p}).
\end{eqnarray}

\par Assume a connection over $\widetilde{P}$ with some element of $\widetilde{\widetilde{P}}$, say $\widetilde{\widetilde{p}}_0$, and define $\Pi$ as its pull-back through $i$ \footnote{In fact, as it will be clear, any different selection for $\widetilde{\widetilde{p}}_0$ has no effect in our formalism.}. Particularly, $\Pi$ is a $\widetilde{\frak{g}}$-valued 1-form over $\widetilde{P}\times \widetilde{G}\times \widetilde{\widetilde{G}}$ as;
\begin{eqnarray} \label {5}
\Pi_{(\widetilde{p},\widetilde{g},\widetilde{\widetilde{g}})}=i^*(\widetilde{\widetilde{p}}_0)_{(\widetilde{p},\widetilde{g},\widetilde{\widetilde{g}})}=Ad_{\widetilde{\widetilde{g}} (\widetilde{x})^{-1}\widetilde{g}^{-1}} (\widetilde{\widetilde{p}}_0)_{\widetilde{p}}+Ad_{\widetilde{\widetilde{g}}(\widetilde{x})^{-1}}(\widetilde{\omega}_{\widetilde{g}})+\widetilde{\widetilde{\omega}}_{\widetilde{\widetilde{g}}(\widetilde{x})},
\end{eqnarray}

\noindent for $\widetilde{\omega}$ and $\widetilde{\widetilde{\omega}}$ the standard Cartan forms on $\widetilde{G}$ and $\widetilde{\widetilde{G}}$ respectively. Using $\Pi$ we define a $\frak{g}$-valued 1-form over $P\times\widetilde{P}\times \widetilde{G}\times \widetilde{\widetilde{G}}$, say $\Omega$, with
\begin{eqnarray} \label {6}
\Omega((v,\widetilde{v},\widetilde{\alpha},\widetilde{\widetilde{\alpha}})_{(p,\widetilde{p},\widetilde{g},\widetilde{\widetilde{g}})})=\widetilde{p}.\widetilde{g} \widetilde{\widetilde{g}} (\widetilde{x})(v_p)+\Pi((\widetilde{v},\widetilde{\alpha},\widetilde{\widetilde{\alpha}})_{(\widetilde{p},\widetilde{g},\widetilde{\widetilde{g}})})(x),
\end{eqnarray}

\noindent for $\widetilde{x}=\widetilde{\pi}(\widetilde{p})$ and $x=\pi (p)$.\\

\par To apply these structures to Yang-Mills gauge theories we need some more assumptions. In fact, any section of $(\widetilde{P},\widetilde{\pi},\widetilde{X})$, if exists, is a gauge fixing term, the mechanism of which selects a single element among any collection of gauge fields which differ one by one only in some gauge transformation. However, as mentioned above $(\widetilde{P},\widetilde{\pi},\widetilde{X})$ may not admit any global section in general. But for simplicity and to get more intuitions one can momentarily assume the triviality of $(\widetilde{P},\widetilde{\pi},\widetilde{X})$. In such a case, this principal bundle admits a global section, say $\widetilde{s}:\widetilde{X}\to \widetilde{P}$, and any other such section, say $\widetilde{s}'$, can be considered as $\widetilde{s}.\widetilde{\widetilde{g}}$ for some appropriate $\widetilde{\widetilde{g}}\in \widetilde{\widetilde{G}}$. Since each of such sections represents a gauge fixing mechanism, an element of $\widetilde{\widetilde{G}}$, actually, changes a gauge fixing term into another one. In fact, this is the reason of which we chose our terminology for such elements, i.e. gauge fixing transformations. Therefore, it must be clear that the gauge fixing transformation group $\widetilde{\widetilde{G}}$ should be a symmetry group for a quantum gauge theory, the fact of which will be explained more in the following.\\

\par If $(\widetilde{P},\widetilde{\pi},\widetilde{X})$ has a non-trivial topology one gets more complications. However, to model Yang-Mills gauge theories in a consistent manner we require some global structures in order to formulate the path-integral formalism with respect to Faddeev-Popov quantization approach. Thus, we may apply some strict  assumptions on $(\widetilde{P},\widetilde{\pi},\widetilde{X})$ to get suitable geometric properties for building up gauge fixing terms. In fact, to overcome the Gribov ambiguity one may suggest different methods. The simplest one, as one consider, is to devide $\widetilde{X}$ into open patches each on which $(\widetilde{P},\widetilde{\pi},\widetilde{X})$ is trivial. Hence, according to this assumption, specifying one of such trivializing open sets, say $U \in \widetilde{X}$, we consider Cartan connection forms over $P$ which belong to $\widetilde{\pi}^{-1}(U)$. Using an open covering of such trivializing patches and assuming a partition of unity subordinated to this covering one can apply the mentioned gauge fixing formalism accordingly to path-integral formulation of quantum Yang-Mills theories \cite{Zwanziger, Klenhofer}.\\

\noindent Now, let $\Gamma_U(\widetilde{\pi})$ be the collection of smooth local sections of $(\widetilde{P},\widetilde{\pi},\widetilde{X})$ which are defined over $U$. Particularly, any element of $\Gamma_U(\widetilde{\pi})$ is a local gauge fixing term and any two of such elements could be correlated with an action of some gauge fixing transformation $\widetilde{\widetilde{g}}\in \widetilde{\widetilde{G}}$. Moreover, if $U \cap U'  \ne \emptyset $, then for any two elements of $\Gamma_U(\widetilde{\pi})$ and $\Gamma_{U'}(\widetilde{\pi})$, say respectively $s$ and $s'$, $s$ and $s'$ differ only in some gauge transformation over $U \cap U'$. Hence, by attaching such elements in compatible manner due to gauge transformations one can overcome the Gribov ambiguity in path-integral formalism \cite{Zwanziger, Klenhofer}. \\

\par Here, for simplicity and without losing any generality let us assume the triviality of $(\widetilde{P},\widetilde{\pi},\widetilde{X})$ \footnote{Unless we should restrict ourselves to one of trivializing patches of $(\widetilde{P},\widetilde{\pi},\widetilde{X})$, say $U$, while $\widetilde{\widetilde{G}}$ then consists of smooth functions $\widetilde{\widetilde{g}}:U\to \widetilde{G}$.}. Thus, fix $\widetilde{p}_0\in \widetilde{P}$ and define for given sections $s\in \Gamma(\pi)$ and $\widetilde{s}\in \Gamma(\widetilde{\pi})$ (the gauge fixing term) the lifting map $j$ with;
\begin{eqnarray} \label {7}
\begin{array}{*{20}{c}}
 ~~~~~~~~  j:X\times \widetilde{G}\times \widetilde{\widetilde{G}}\to P\times\widetilde{P}\times \widetilde{G}\times \widetilde{\widetilde{G}}~~~~~~~~~~~~~~~  \\
   ~~(x,\widetilde{g},\widetilde{\widetilde{g}})~~~\mapsto (s(x),\widetilde{s}(\widetilde{x}_0)),\widetilde{g},\widetilde{\widetilde{g}}).  \\
\end{array}
\end{eqnarray}

\noindent for $\widetilde{x}_0=\widetilde{\pi}(\widetilde{p}_0)$. The pull-back of $\Omega$ through $j$ is a $\frak{g}$-valued 1-form over $X\times \widetilde{G}\times \widetilde{\widetilde{G}}$ as;
\begin{eqnarray} \label {8}
j^*(\Omega)_{(x,\widetilde{g},\widetilde{\widetilde{g}})}=s^*(\widetilde{s}(\widetilde{x}_0).\widetilde{g}\widetilde{\widetilde{g}}(\widetilde{x}_0))_x+Ad_{\widetilde{\widetilde{g}}(\widetilde{x}_0)^{-1}}(\widetilde{\omega}_{\widetilde{g}})(x)+\widetilde{\widetilde{\omega}}_{\widetilde{\widetilde{g}}}(\widetilde{x}_0)(x).
 \end{eqnarray}

\noindent Let's define the "gauge field" $A$ as the first term of $j^*(\Omega)$, i.e.;
\begin{eqnarray} \label {9}
A_{(x,\widetilde{g},\widetilde{\widetilde{g}})}=s^*(\widetilde{s}(\widetilde{x}_0).\widetilde{g}\widetilde{\widetilde{g}}(\widetilde{x}_0))_x.
\end{eqnarray}
 
\noindent Show the second term and the third term of $j^*(\Omega)$ with $C$ and $\widetilde{C}$ and let us refer to them respectively with "ghost" and "anti-ghost" fields. Thus;
\begin{eqnarray} \label {10}
\begin{array}{*{20}{c}}
C_{(x,\widetilde{g},\widetilde{\widetilde{g}})}=Ad_{\widetilde{\widetilde{g}}(\widetilde{x}_0)^{-1}}(\widetilde{\omega}_{\widetilde{g}})(x)  \\
\widetilde{C}_{(x,\widetilde{g},\widetilde{\widetilde{g}})}=\widetilde{\widetilde{\omega}}_{\widetilde{\widetilde{g}}}(\widetilde{x}_0)(x).  ~~~~~~~~ \\
\end{array}
\end{eqnarray}

\par For working out the full BRST-anti-BRST structures we need nilpotent operators for BRST and anti-BRST derivatives. To find these operators note that the exterior derivative operator on $X\times \widetilde{G}\times \widetilde{\widetilde{G}}$ naturally splits to three terms for exterior derivatives on $X$, $\widetilde{G}$ and $\widetilde{\widetilde{G}}$ which are respectively denoted by $\emph{\emph{d}}$, $\delta$, and $\widetilde{\delta}$. Direct calculation shows that;
\begin{eqnarray} \label {11}
\begin{array}{*{20}{c}}
   \delta A=-\emph{\emph{d}}C-AC-CA,  \\
   \delta C=-C^2,~~~~~~~~~~~~~~~~~  \\
   \delta \widetilde{C}=0,~~~~~~~~~~~~~~~~~~~~~  \\
\end{array}
\end{eqnarray}

\noindent and;
\begin{eqnarray} \label {12}
\begin{array}{*{20}{c}}
   \widetilde{\delta} A=-\emph{\emph{d}}\widetilde{C}-A\widetilde{C}-\widetilde{C}A,  \\
   \widetilde{\delta} C=-C\widetilde{C}-\widetilde{C}C,~~~~~~~~  \\
  \widetilde{\delta} \widetilde{C}=-\widetilde{C}^2.~~~~~~~~~~~~~~~~~  \\
\end{array}
\end{eqnarray}

\noindent On the other hand, since $\delta$ and $\widetilde{\delta}$ are exterior derivatives on $\widetilde{G}$ and $\widetilde{\widetilde{G}}$ we directly read;
\begin{eqnarray} \label {13}
\delta^2=\widetilde{\delta}^2=0,
\end{eqnarray}

\noindent while since $\emph{\emph{d}}+\delta+\widetilde{\delta}$ is itself a nilpotent operator as the exterior derivative operator on $X\times \widetilde{G}\times \widetilde{\widetilde{G}}$ we conclude;
\begin{eqnarray} \label {14}
\emph{\emph{d}}\delta+\delta\emph{\emph{d}}=\emph{\emph{d}}\widetilde{\delta}+\widetilde{\delta}\emph{\emph{d}}=\delta\widetilde{\delta}+\widetilde{\delta}\delta=0.
\end{eqnarray}

\noindent Setting the auxiliary Nakanishi-Lautrup field equal to zero\footnote{Which is the case in strict gauge fixing mechanisms such as that of Landau.}, one finds that the relations (\ref {11}) to (\ref {14}) are in complete agreement with those of BRST and anti-BRST derivatives if one show them with $\delta$ and $\widetilde{\delta}$ respectively. Therefore, let us refer to $\delta$ and $\widetilde{\delta}$ with "BRST" (derivative/operator) and "anti-BRST" (derivative/operator) accordingly.\\

\par To write down the above structures in a more familiar setting we should note that $j^*(\Omega)=A+C+\widetilde{C}$ can naturally be considered as a $\frak{g}$-valued 1-form over $X\times \widetilde{X} \times \widetilde{G} \times \widetilde{\widetilde{G}}$ as;
\begin{eqnarray} \label {15}
\begin{array}{*{20}{c}}
   A_{(x,\widetilde{x},\widetilde{g},\widetilde{\widetilde{g}})}=s^*(\widetilde{s}(\widetilde{x}).\widetilde{g}\widetilde{\widetilde{g}}(\widetilde{x}))_x,  \\
  C_{(x,\widetilde{x},\widetilde{g},\widetilde{\widetilde{g}})}=Ad_{\widetilde{\widetilde{g}}(\widetilde{x})^{-1}}(\widetilde{\omega}_{\widetilde{g}})(x),  \\
  \widetilde{C}_{(x,\widetilde{x},\widetilde{g},\widetilde{\widetilde{g}})}=\widetilde{\widetilde{\omega}}_{\widetilde{\widetilde{g}}}(\widetilde{x})(x).~~~~~~~~~   \\
\end{array}
\end{eqnarray}

\noindent In this setting the differential forms of $A$, $C$, and $\widetilde{C}$, and consequently the Faddeev-Popov quantum Lagrangian of a gauge theory in path-integral formalism, all are considered as some geometric objects over the space of $X \times \widetilde{P} \times \widetilde{\widetilde{G}}$ \footnote{This simplification is due to considering the triviality of $\widetilde{P}$, i.e. $\widetilde{P}=\widetilde{X}\times \widetilde{G}$.} which beside its third component is the familiar space that we path-integrate over to quantize the theory. Note that in this framework the only fixed element (beside $s$ \footnote{The term of which is trivially defined for the contractable space-time $X=\mathbb{R}^4$ as $s=1$, for $1$ the identity elelment of $G$. }) is the gauge fixing term $\widetilde{s}$ which we need as a fundamental tool for Faddeev-Popov quantization method of (non-Abelian) Yang-Mills theories. \\

\par Appearing $\widetilde{\widetilde{G}}$ in defining relations of (\ref {15}) is based on the varieties of different gauge fixing mechanisms. Therefore, it would be the most expected property of a gauge theory which has no reference to gauge fixings in its classical formalism to be symmetric under the action of $\widetilde{\widetilde{G}}$ at quantum levels. More precisely, the Faddeev-Popov quantum Lagrangian as a differential form on $X \times \widetilde{P} \times \widetilde{\widetilde{G}}$ has to be a closed form for derivative operator $\widetilde{\delta}$. This is the simplest geometric explanation of anti-BRST invariance which was reported for the first time as an  accidental symmetry of Faddeev-Popov quantum Lagrangian \cite {Henneaux1, Weinberg, Alvarez-Gaume}.\\

%%%%%%%%%%%%%%%%%%%%%%%%%%%%%%%%%%%%%%%%%%%%%%%%%%%%%

\par
\section{More about anti-BRST Structure}
\setcounter{equation}{0}

\par For the first attempt we try to provide a simple brief proof for validity of our model in the previous section. In order to do this we should remind that for any physically consistent gauge theory at classical level the Lagrangian density never addresses to gauge fixing methods. Consequently, gauge fixing invariance ceases to provide an independent symmetry beside the gauge invariance in such theories. However, when we quantize classical gauge theories via the Faddeev-Popov mechanism, the gauge fixing function, say $f$, automatically enters into the quantum Lagrangian (via a gauge fixing term, e.g. $\frak{L}_{GF}=\frac{1}{2}~tr\{f^2\}+...~$). But since $f$ restricts the domain of path-integration to its roots (i.e. gauge field $A$ with $f(A)=0$) the theory selects a unique gauge field for each equivalence class of gauge fields\footnote{specially it is the case for theories with no Gribov ambiguity such as Abelian ones.} (or any fiber of $\widetilde{P}$) within path-integral formulation\footnote{Here one can focus on sharp gauge fixings for more relevant intuition.}. Therefore, any continuous transformation in $f$, such as $f \to f'=f+\delta f$, for $|\delta f|  \ll 1$, leads naturally to an infinitesimal gauge transformation, say $\alpha$, to $A$, i.e. $A \to A'= A+A.\alpha=A+\emph{\emph{d}}\alpha+[A,\alpha]$, to make $A'$ a root of new gauge fixing function $f'$ and incorporate it into the path-integral. Considering $\alpha$ as infinitesimal gauge fixing transformation and $\widetilde{\delta}$ as its relevant transformation to quantum fields, we naturally find out;
\begin{eqnarray} \label {3-1}
 \widetilde{\delta}A=-\emph{\emph{d}}\widetilde{C}-A\widetilde{C}-\widetilde{C}A,
\end{eqnarray}

\noindent where $\widetilde{C}$ stands for a differential 1-form dual to $\alpha$. Moreover, a subsequent infinitesimal gauge fixing transformation, say $\beta$, leads to infinitesimal gauge fixing tansformation  $[\alpha,\beta]$. In fact, this is also the case for infinitesimal gauge fixing transformation $\alpha$ followed by infinitesimal gauge transformation $\beta$ which similarly results in $[\beta,\alpha]$ as the consequent infinitesimal gauge transformation. Actually, both these cases are simply given by Maurer-Cartan formula as;
\begin{eqnarray} \label {3-2}
 \widetilde{\delta}\widetilde{C}=-\widetilde{C}^2,
\end{eqnarray}

\noindent and;
\begin{eqnarray} \label {3-3}
 \widetilde{\delta}C=-C\widetilde{C}-\widetilde{C}C.
\end{eqnarray}

\noindent The relations of (\ref{3-1})-(\ref{3-3}) are actually those of well-known anti-BRST transformations\footnote{with no Nakanashi-Lautrup field.} just as we derived in (\ref{12}). This is the result of which we aimed to prove here, although not with strict mathematics, to confirm our elaborated geometric model for anti-BRST structure in the previous section.\\

\noindent Here, due to the physical consistency in classical version of the theory we expect the quantized theory to be invariant under variation of this gauge fixing function. In fact, as we showed above, the invariance of quantized gauge theory under anti-BRST transformations must follow naturally. This consequence makes the anti-BRST invariance indispensable for BRST symmetric quantum gauge theories, the fact of which will be explained with more details in the following. \\

\par Now we prefer to have a brief discussion about independence of gauge and gauge fixing symmetries in their own general formalisms. In order to see this it is enough to define a gauge symmetric functional, say $\frak{f}$, which is not gauge fixing invariant, and to introduce a gauge fixing invariant functional, say $\frak{h}$, which is not gauge symmetric. For simplicity let's restrict ourselves to trivial fibration $(\widetilde{P},\widetilde{\pi},\widetilde{X})$. Therefore, any gauge fixing function, say $s$, would be a global section of $\widetilde{\pi}:\widetilde{P} \to \widetilde{X}$. Consider a parallelism structure over this principal bundle and decompose $T\widetilde{P}$ to $T\widetilde{X} \oplus \widetilde{\frak{g}}$ accordingly. Then, restricting the image of $\emph{\emph{d}}s:T\widetilde{X} \to T\widetilde{P}$ to $\widetilde{\frak{g}}$ provides a $\widetilde{\frak{g}}$-valued 1-form over $\widetilde{X}$. We show this restriction with $\emph{\emph{d}}_Vs$. Thus, for any fixed $\widetilde{\frak{g}}$-valued vector field over $\widetilde{X}$, say $Y$, 
\begin{eqnarray} \label {3-0}
\begin{array}{*{20}{c}}
   \frak{f}: \widetilde{P} \times \Gamma(\widetilde{\pi})~~ \to ~~C^{\infty}_0(X)  \\
   ~~~~~~~~~~~~~~~~~~~(\widetilde{p},s) ~~~~~~\mapsto ~~tr\{\emph{\emph{d}}_Vs(Y)\}(\widetilde{\pi}(\widetilde{p})),  \\
\end{array}
\end{eqnarray}

\noindent provides a gauge invariant functional over $X$, where here $tr$ stands for taking trace in $\frak{g}$ \footnote{Recall that $\widetilde{\frak{g}}=\frak{g} \otimes_{\mathbb{C}} C_0^{\infty} (X)$. Thus, $tr$ acts only on the first component.}. Since $\frak{f}$ depends on behavior of $s$ near $\widetilde{\pi}(\widetilde{p})$ we eventually conclude that it is not a gauge fixing symmetric functional. On the other hand, $\frak{f}$ is a function of $\widetilde{\pi}(\widetilde{p})$ but not $\widetilde{p}$, thus obviously it is gauge symmetric. Roughly speaking we find 
\begin{eqnarray} \label {3-000}
\frac{{\emph{\emph{d}}\frak{f}}}{{\emph{\emph{d}}s}} \ne 0  ~~~~~\emph{\emph{and}}~~~~~ \frac{{\emph{\emph{d}}\frak{f}}}{{\emph{\emph{d}}\widetilde{g}}} = 0.
\end{eqnarray}

\noindent This proves that gauge fixing symmetry is independent of gauge invariance in general. In other words, due to our model for BRST-anti-BRST formalism we read from (\ref{3-000});
\begin{eqnarray} \label {3-001}
\widetilde{\delta}\frak{f} \ne 0  ~~~~~\emph{\emph{and}}~~~~~ \delta \frak{f}= 0.
\end{eqnarray}

\noindent Conversely, to see the independence of gauge invariance form gauge fixing symmetry one should note that any function $\frak{h}: \widetilde{P} \times \Gamma(\widetilde{\pi}) \to C^{\infty}_0(X)$ which varies through with gauge transformations and gives no address to gauge fixing $s$ is in its own nature a gauge fixing invariant functional over $X$ that is not gauge symmetric\footnote{Thus, $\frak{h}$ is essentially given as $\frak{h}:\widetilde{P} \to C^{\infty}_0(X)$.}. Thus, similarly we find;
\begin{eqnarray} \label {3-001}
\frac{{\emph{\emph{d}}\frak{h}}}{{\emph{\emph{d}}s}} = 0  ~~\emph{\emph{and}}~~ \frac{{\emph{\emph{d}}\frak{h}}}{{\emph{\emph{d}}\widetilde{g}}} \ne 0 ~~~~~\emph{\emph{or equivalently;}}~~~~~  \widetilde{\delta}\frak{h} = 0  ~~\emph{\emph{and}}~~ \delta \frak{h} \ne 0.
\end{eqnarray}

\noindent These results show that as gauge invariance and gauge fixing symmetry are two independent structures at classical levels, BRST and anti-BRST symmetries are two independent quantum features. Hence, due to our model two independent classical symmetries of gauge and gauge fixing invariance lead to two independent quantum symmetries of BRST and anti-BRST accordingly. \\

\noindent However, in standard Yang-Mills theories which the gauge symmetric Lagrangian density doesn't refer to gauge fixing $s$ the theory admits both these two classical symmetries automatically and we interpret this situation as physical consistency. More precisely, in such cases $\frak{L}_{Y-M}$ is a map from $\widetilde{P}$ to $C^{\infty}_0(X)$ which could be redefined with
\begin{eqnarray} \label{3-00}
\frak{L}_{Y-M}=\frak{L}_{Y-M} \circ s \circ \widetilde{\pi}
\end{eqnarray}

\noindent  for any gauge fixing $s$. This actually defines a map over $\widetilde{P} \times \Gamma(\widetilde{\pi})$ (to $C^{\infty}_0(X)$) on which we can evaluate the derivatives $\frac{{\emph{\emph{d}}}}{{\emph{\emph{d}}\widetilde{g}}}$ and $\frac{{\emph{\emph{d}}}}{{\emph{\emph{d}}s}}$. Appearing $\widetilde{\pi}$ on the right hand side of (\ref{3-00}) ensures gauge invariance (i.e. $\frac{{\emph{\emph{d}}{\frak{L}_{Y-M}}}}{{\emph{\emph{d}}\widetilde{g}}}=0$) and disappearing $s$ on the left hand side guaranties gauge fixing symmetry (i.e. $\frac{{\emph{\emph{d}}{\frak{L}_{Y-M}}}}{{\emph{\emph{d}}s}}=0$). Now one can follow our establishments in previous section to study the quantum features of these two classical symmetries, the procedure which leads to BRST and anti-BRST invariance in quantized gauge theories. More precisely, in physically consistent gauge theories BRST and anti-BRST symmetries both are indispensable for quantum Lagrangian density $\frak{L}$, i.e.;
\begin{eqnarray} \label {3-002}
\widetilde{\delta}\frak{L} = 0  ~~~~~\emph{\emph{and}}~~~~~ \delta \frak{L}= 0.
\end{eqnarray}

\noindent Therefore, actually it may be concluded that BRST and anti-BRST symmetries are not independent for consistent gauge theories in general off-shell formalisms \cite{Henneaux5, Upadhyay, Blagojevic}. This conclusion, however, explains the mystical accidentally appearance of anti-BRST symmetry in quantum Lagrangian of Yang-Mills theories \cite{Alvarez-Gaume}.

%%%%%%%%%%%%%%%%%%%%%%%%%%%%%%%%%%%%%%%%%%%%%%%%%%%%%%%%

\par
\section{Topology and Geometry of anti-BRST Structure}
\setcounter{equation}{0}

\par Now, by discovering a compatible background model for anti-BRST structure the fascinating features of the topology and the geometry of anti-BRST invariance could be investigated with more details. As we noted above anti-BRST invariance is an inevitable symmetry together with the BRST one for quantized gauge theories. In the following by studying the topology and the geometry of anti-BRST more evidences for agreements between BRST and anti-BRST structures in quantized gauge theories are derived. \\

\par One can use the method of consistent anomalies via the external field formalism to study the structure of anti-BRST due to our given model. It is also possible to reestablish the coincidence of BRST and anti-BRST symmetries in quantized gauge theories through with the same method. To see this more strictly recall that the consistent anomaly is given as the BRST derivation of the effective action in presence of an external field as the source. In fact, if $W(A)$ is the effective action of a quantum Yang-Mills theory for external field $A$, i.e.;
\begin{eqnarray}  \label {2-1}
e^{iW(A)}=\int {D\psi D\overline \psi ~e^{iS(\psi,\overline{\psi},A)} },
\end{eqnarray}

\noindent then, the consistent anomaly is defined as;
\begin{eqnarray}  \label {2-2}
\frak{A}(A):=\delta W(A)=\int_{\mathbb{R}^4}~C^a (\emph{\emph{D}}J)^a~ \emph{\emph{d}}^4x,
\end{eqnarray}

\noindent where $(\emph{\emph{D}}J)^a=\partial_\mu J^{a~\mu}+f^{abc}A^b_\mu J^{c~\mu}$, for $J^{a~\mu}=\overline{\psi}\gamma^\mu t^a \psi$ the current of matter field, and $f^{abc}$ the structure constant of semi-simple gauge group $G$ for Lie algebra basis $\{ t^a\}_{a=1}^{\emph{\emph{dim}} G}$. Here $C^a$ is that component of ghost field $C$ which couples with $t^a$. That is $\frak{A}(A)$ defines a differential 1-form over $\widetilde{G}$ (or $\widetilde{P}$). Then, for any infinitesimal gauge transformation $\widetilde{\alpha}\in \widetilde{\frak{g}}$ the evaluated anomaly is given as;
\begin{eqnarray}  \label {2-2-1}
\frak{A}(A)(\widetilde{\alpha})=\int_{\mathbb{R}^4}~\widetilde{\alpha}^a (\emph{\emph{D}}J)^a~ \emph{\emph{d}}^4x.
\end{eqnarray}

\noindent Hence, $\frak{A}(A)=0$ if and only if the quantum theory is consistently anomaly free, i.e. $\frak{A}(A)=0$ if and only if $(\emph{\emph{D}}J)^a=0$ for all indices $a$.\\

\par On the other hand, similarly we define
\begin{eqnarray}  \label {2-3}
\widetilde{\frak{A}}(A):=\widetilde{\delta} W(A)=\int_{\mathbb{R}^4}~\widetilde{C}^a (\emph{\emph{D}}J)^a~ \emph{\emph{d}}^4x
\end{eqnarray}

\noindent as a differential 1-form over $\widetilde{\widetilde{G}}$. Let's refer to $\frak{A}(A)$ and $\widetilde{\frak{A}}(A)$ with ghost consistent anomaly and anti-ghost consistent anomaly respectively. In fact, due to the formulation of $\widetilde{\frak{A}}(A)$ we conclude as the first result that the anti-ghost consistent anomaly vanishes if and only if the gauge invariance is preserved after quantization in presence of external field $A$.\\

\noindent One can establish this achievement through with some different ways. To see this note that applying descent equations method starting with six-dimensional Chern character via the index theorem one precisely shows that ghost/anti-ghost consistent anomalies of $(3+1)$-dimensional Yang-Mills theory are given as \cite{Weinberg};
\begin{eqnarray}  \label {2-3-1}
   {\frak{A}(A)=\frac{{1}}{{24 \pi^2}}~\int_{\mathbb{R}^4}~tr \{ \emph{\emph{d}}C (A\emph{\emph{d}}A+\frac{{1}}{{2}}A^3) \}},
\end{eqnarray}

\noindent and
\begin{eqnarray}  \label {2-3-2}
   {\widetilde{\frak{A}}(A)=\frac{{1}}{{24 \pi^2}}~\int_{\mathbb{R}^4}~tr \{ \emph{\emph{d}}\widetilde{C} (A\emph{\emph{d}}A+\frac{{1}}{{2}}A^3) \}},
\end{eqnarray}

\noindent respectively. In fact, the same similarity holds between ghost and anti-ghost consistent anomalies in other dimensions, i.e. there is a fixed functional, say $\frak{F}$, such that;
\begin{eqnarray}  \label {2-33}
\frak{A}(A)=\frak{F}(A;C)~~~~~\emph{\emph{and}}~~~~~\widetilde{\frak{A}}(A):=\frak{F}(A;\widetilde{C}).
\end{eqnarray}

\noindent Actually, we again find;
\begin{eqnarray} \label {2-3-3}
\frak{A}(A)=0~~~~\emph{\emph{if and only if}}~~~~\widetilde{\frak{A}}(A)=0. 
\end{eqnarray}

\noindent This result confirms our proof for BRST symmetry-anti-BRST symmetry agreement in quantized Yang-Mills theories. Particularly, (\ref{2-3-3}) states that once either BRST or anti-BRST symmetry is lost, the other symmetry will not be preserved. In other words, since nonvanishing $\frak{A}(A)$ and $\widetilde{\frak{A}}(A)$ are respectively the criterions for disappearing gauge and gauge fixing invariances after quantization, we conclude from (\ref{2-3-3}) whenever the former is anomalously broken the later breaks automatically and vice versa.\\

\par However, to have more complete details about topological meanings of $\frak{A}(A)$ and $\widetilde{\frak{A}}(A)$ one should focus on cohomology classes $[\frak{A}(A)]$ and $[\widetilde{\frak{A}}(A)]$ in $H^1_{\emph{\emph{de~Rham}}}(\widetilde{G},\mathbb{R})$ and $H^1_{\emph{\emph{de~Rham}}}(\widetilde{\widetilde{G}},\mathbb{R})$ respectively. In fact, $W(A)$ is a non-local term in general, then its BRST and anti-BRST derivatives, shown accordingly with $\frak{A}(A)$ and $\widetilde{\frak{A}}(A)$, may belong to non-trivial cohomology classes generaly. Although using appropriate counter terms via renormalization methods will change the quantum action $W(A)$ to $W'(A)=W(A)+c(A)$ for counter term $c(A)$ \footnote{Counter term $c(A)$ either may cause the quantum gauge theory to preserve gauge invariance in price of missing the axial symmetry at quantum levels or it may be used to give simple structures to Green's functions of the theory as noted in the introduction.} \cite{Bertlmann} (and consequently will replace $\frak{A}(A)$ and $\widetilde{\frak{A}}(A)$ with $\frak{A}'(A):=\delta W'(A)$ and $\widetilde{\frak{A}}'(A):=\widetilde{\delta} W'(A)$), this will not affect the cohomology classes $[\frak{A}(A)]$ and $[\widetilde{\frak{A}}(A)]$\footnote{Since $c(A)$ is a smooth function its BRST/anti-BRST derivaties are BRST/anti-BRST exact forms.}. Therefore, (\ref{2-3-3}) takes the following topological form;
\begin{eqnarray} \label {2-3-4}
[\frak{A}(A)]=0~~~~\emph{\emph{if and only if}}~~~~[\widetilde{\frak{A}}(A)]=0. 
\end{eqnarray}

\noindent This assertion explains that after quantizing a classical Yang-Mills gauge theory with background field $A$ the gauge invariance is lost topologically if and only if gauge fixing symmetry is missed with the same meaning. More precisely, just as gauge and gauge fixing invariances are mutually inevitable in classical formalisms, topologically the triviality of BRST structure at quantum levels leads to the triviality of that of anti-BRST and vice versa. Thus, anti-BRST symmetry (which was initially understood as an accidentaly mysterious emergent symmetry of quantum Lagrangian \cite{Henneaux1}) is topologically indispensable for a BRST invariant quantized gauge theory. \\

\par To clear our meaning of BRST-anti-BRST topological structures and to see more of such agreements between these settings we need to shed more light on geometric-topological properties of ghost and anti-ghost consistent anomalies. Essentially, according to family index theorem $\frak{A}(A)$ is the Cartan connection form for the norm-invariant parallelism structure of the deteminant line bundle of solutions of left and right handed Dirac operators over $\widetilde{P}$ \cite{Bismut1, Bismut2}. However, $\frak{A}(A)$ can be pulled back to $\widetilde{G}$ as a fixed fiber of $\widetilde{P}$. We show this determinant line bundle with $\widehat{DET} \to \widetilde{G}$ or simply with $\widehat{DET}$. Since $\delta$ is nilpotent, $\delta^2=0$, one concludes the celebrated Wess-Zumino consistency condition $\delta \frak{A}(A)=0$, which consequently leads to flatness of $\widehat{DET}$. This also corresponds a de Rham cohomology class to ghost anomaly $\frak{A}(A)$. \\

\noindent Similarly the same is true for anti-ghost consistent anomaly. In order to show this one should note that by our constructions the anti-ghost field is the pull-back of the ghost through $\phi_{\widetilde{x}}$ for some $\widetilde{x} \in \widetilde{X}$, i.e. $\widetilde{C}=\phi_{\widetilde{x}}^*(C)$. In fact, due to (\ref{2-33}) $\widetilde{\frak{A}}(A)$ will be the pull-back of $\frak{A}(A)$ through $\phi_{\widetilde{x}}$, i.e. $\widetilde{\frak{A}}(A)=\phi_{\widetilde{x}}^*(\frak{A}(A))$. More precisely, $\widetilde{\frak{A}}(A)$ is the connection form of a line bundle over $\widetilde{\widetilde{G}}$,  say $\widetilde{DET}$, which is topologically the pull back of $\widehat{DET}$ through $\phi_{\widetilde{x}}$. In other words, the line bundle $\widetilde{DET}$ is definitely equal to $\phi_{\widetilde{x}}^*(\widehat{DET})$ and its parallelism structure is induced from that of $\widehat{DET}$. Therefore, geometrically we conclude that $\frak{A}(A)$ is flat if and only if $\widetilde{\frak{A}}(A)$ is flat too, the fact of which could also be deduced from $\widetilde{\delta}^2=0$. This results in an anti-BRST version of Wess-Zumino consistency condition; $\widetilde{\delta} \widetilde{\frak{A}}(A)=0$, the relation which geometrically is understood as the flatness of $\widetilde{DET}$. Thus, the quantized theory is BRST symmetric if and only if it is anti-BRST invariant. \\

\par The flatness of $\frak{A}(A)$ leads topologicaly to triviality of $\widehat{DET}$ whenever $\widetilde{G}$ is simply-connected. Actually, this is also  the case for $\widetilde{DET}$ over $\widetilde{\widetilde{G}}$, i.e. for simply-connected $\widetilde{\widetilde{G}}$ the flatness of $\widetilde{\frak{A}}(A)$ leads to trivial topology of $\widetilde{DET}$. But for Yang-Mills theories with general compact gauge groups $G$, the gauge transformation group $\widetilde{G}$ (and consequently the gauge fixing transformation group $\widetilde{\widetilde{G}}$) is not simply-connected, then the triviality of  topology of $\widehat{DET}$ and $\widetilde{DET}$ could not be infered accordingly. However, although in such cases we usually need more complicated analysis, similar results about the same topology of $\widehat{DET}$ and $\widetilde{DET}$ follows simply as we will prove in below. \\

\par Actually, due to family index theorem $[\frak{A}(A)]$ and $[\widetilde{\frak{A}}(A)]$ belong to integral cohomology classes of $H^1(\widetilde{G},\mathbb{Z})$ and $H^1(\widetilde{\widetilde{G}},\mathbb{Z})$ respectively \cite{Bismut1, Bismut2}. That is for any one-dimensional inclusion $i:S^1 \to \widetilde{G}$, we find \cite{Weinberg, Bertlmann, Bismut1};
\begin{eqnarray} \label {2-3-5}
\int_{S^1}~i^*(\frak{A}(A))~\in~\mathbb{Z}. 
\end{eqnarray}

\noindent More precisely, the parallelism structure of $\widehat{DET}$ provides a finite covering space for $\widetilde{G}$, say $\widetilde{\frak{G}}$, and any holonomy element of $\widehat{DET}$ refers to an inverse image in this covering. Hence, the index of $\frak{A}(A)$ against $i:S^1 \to \widetilde{G}$ counts the winding number of $i$ in $\widetilde{\frak{G}}$ which is in fact the phase of the corresponding holonomy element of $\widehat{DET}$ modulo some factor of $2\pi$. However, it is clear that for any smooth $i:S^1 \to \widetilde{G} \subset \widetilde{\widetilde{G}}$ we calculate
\begin{eqnarray} \label {2-3-6}
\int_{S^1}~i^*(\frak{A}(A))=\int_{S^1}~i^*(\widetilde{\frak{A}}(A))~\in~\mathbb{Z}. 
\end{eqnarray}

\noindent In fact, if $\phi_{\widetilde{x}}$ and the natural inclusion of $\widetilde{G} \subset \widetilde{\widetilde{G}}$ provide a deformation retract, then $\phi^*_{\widetilde{x}}$ gives rise to isomorphisms between $H^*(\widetilde{G},\mathbb{Z})$ and $H^*(\widetilde{\widetilde{G}},\mathbb{Z})$, the case in which (\ref {2-3-6}) concludes naturally. But, otherwise, when $\widetilde{G}$ and $\widetilde{\widetilde{G}}$ are not homotopic equivalent, and consequently we find generally $H^*(\widetilde{G},\mathbb{Z}) \ne H^*(\widetilde{\widetilde{G}},\mathbb{Z})$, equation (\ref {2-3-6}) leads to a non-trivial result for cohomology classes $[\frak{A}(A)]$ and $[\widetilde{\frak{A}}(A)]$. Consequently, according to index theorem BRST and anti-BRST settings provide the same topological structures over $\widetilde{G}$ and $\widetilde{\widetilde{G}}$ via the mentioned line bundles\footnote{In fact, $\widehat{DET}$ is also the pull-back of $\widetilde{DET}$ through natural embedding of $\widetilde{G} \subset \widetilde{\widetilde{G}}$}. Therefore, due to (\ref{2-3-6}) any continuous deformation of gauge theory due to counter term $c(A)$, which may cause some new effects in the theory against BRST and anti-BRST derivations will not affect the coincidence of BRST and anti-BRST indexes. In other words, (\ref{2-3-6}) strictly proves the agreement between the index of BRST and that of anti-BRST in quantized Yang-Mills gauge theories. This reestablishes our previous result with a different statement as if gauge invariance is broken after quantization (i.e. $\int_{S^1}~i^*(\frak{A}(A)) \ne 0$ for some $i:S^1 \to \widetilde{G}$), then the anti-BRST symmetry of the quantized theory is also lost, i.e. $[\widetilde{\frak{A}}(A)] \ne 0$.\\

\par Other usefull results could be deduced by exploring the BRST and anti-BRST cohomologies of physical states. In fact, here we prefer to focus more on the cohomology of BRST charge in the total Hilbert space of states \cite{Weinberg} rather than local BRST cohomology of local forms (functionals). The local BRST cohomology of the BRST differential and the master equation in antifield-antibracket formalism should be studied separately in details\footnote{However, replacing $\delta$ with $\widetilde{\delta}$ in this formalism needs a redefinition of well-known antibracket and consequently of the master equation. This topic will be studied in \cite{Varshovi-next}.}.\\

\noindent To study anti-BRST cohomology of physical states we first need to look at the anti-BRST charge $\widetilde{Q}$. Conventionally, the anti-BRST charge $\widetilde{Q}$, is defined just similar to that of the BRST, $Q$, as an operator over the total Hilbert space of quantum states with;
\begin{eqnarray}  \label {2-4}
[\widetilde{Q},\Phi]=\widetilde{\delta}\Phi~~~\emph{\emph{similar to}}~~~[Q,\Phi]=\delta\Phi,
\end{eqnarray}

\noindent for any operator $\Phi$ including gauge, matter, ghost and anti ghost fields and their combinations. Note that in (\ref {2-4}) the symbol of $[~,~]$ is a super-commutator. According to  relations (\ref {11}) to (\ref {14}), the anti-BRST charge $\widetilde{Q}$, just similar to its BRST counterpart $Q$, must obey the relations of;
\begin{eqnarray} \label {2-5-1}
\begin{array}{*{20}{c}}
   \widetilde{Q}^2=Q^2=0,  \\
   \widetilde{Q}Q+Q\widetilde{Q}=0,  \\
\end{array}
\end{eqnarray}

\noindent and;
\begin{eqnarray} \label {2-5}
  \left\langle \alpha  \right| \widetilde{Q}=\widetilde{Q} \left| \alpha  \right\rangle = \left\langle \alpha  \right| Q=Q \left| \alpha  \right\rangle =0,
\end{eqnarray}

\noindent for any physical state $\left| \alpha  \right\rangle$ \cite{Weinberg, Hull}. In fact, (\ref{2-5}) is a direct consequence of;
\begin{eqnarray} \label {2-6}
   [\widetilde{Q},S]=[Q,S]=0.
\end{eqnarray}

\noindent for $S$ the scattering matrix of the theory.\\

\noindent By definition the Hilbert space of physical states ($Q$ closed elements) modulo BRST images ($Q$ exact elements) is mostly refered to as BRST cohomology of physical states; $H_{\emph{\emph{BRST}}}(Q)$. The cohomology of physical states via the BRST operator $Q$ is the central element that controls the physics \cite{FHM, FHST}. It is seen that these physical states are free of both ghosts and anti-ghosts. Moreover, it can be shown that for general gauge fixing mechanism the $S$-matrix is unitary in this space. However, by appropriately choosing a strict gauge fixing (e.g. the Landau's gauge), as we studied here for vanishing Nakanishi-Lautrup field, one can reduce the path integral to an expression that involves only the physical (transverse) degrees of freedom and which is manifestly unitary in the physical subspace (equal to the reduced phase space path integral) \cite{FHM}. Finally, the negative norm degrees of freedom are also transformed away due to BRST symmetry of the quantized theory \cite{Weinberg}. \\

\par The same argument holds for anti-BRST cohomology of physical states; $H_{\emph{\emph{anti-BRST}}}(\widetilde{Q})$ \footnote{Which is by definition the Hilbert space of physical space ($\widetilde{Q}$ closed elements) modulo anti-BRST images ($\widetilde{Q}$ exact elements).}. In the same way it is seen that the physical states of classes of $H_{\emph{\emph{anti-BRST}}}(\widetilde{Q})$ are cohomologically free of both ghosts and anti-ghosts\footnote{Each elelment is considered modulo ghost and anti-ghost fields.}. Therefore, there is a natural bijection $\zeta:H_{\emph{\emph{BRST}}}(Q) \to H_{\emph{\emph{anti-BRST}}}(\widetilde{Q})$ which sends any class of $H_{\emph{\emph{BRST}}}(Q)$, say $[\left| \alpha  \right\rangle]$, to that class of $H_{\emph{\emph{anti-BRST}}}(\widetilde{Q})$ which contains the ghost-anti-ghost free element of $[\left| \alpha  \right\rangle]$ and vice versa. In particular, one finds;
\begin{eqnarray} \label {2-7}
 H_{\emph{\emph{BRST}}}(Q) \cong H_{\emph{\emph{anti-BRST}}}(\widetilde{Q}).
\end{eqnarray}
 
 \noindent This isomorphism again addresses to similar topological structures of BRST and anti-BRST charges. Moreover, due to our argument above $H_{\emph{\emph{anti-BRST}}}(\widetilde{Q})$ must be also free of negative norm states. On the other hand, by our definition of isomorphism $\zeta$ it would be clear that $S$-matrix is unitary in $H_{\emph{\emph{anti-BRST}}}(\widetilde{Q})$ if and only if it is unitary in $H_{\emph{\emph{BRST}}}(Q)$\footnote{Define the scattering matrix on $H_{\emph{\emph{anti-BRST}}}(\widetilde{Q})$ with $S_{\widetilde{Q}}=\zeta S_Q\zeta^{-1}$ for $S_Q$ the scattering matrix on $H_{\emph{\emph{BRST}}}(Q)$.}. In fact, (\ref{2-7}) claims that there is no difference to characterize the physical states of the theory either by means of BRST classification or by that of anti-BRST. In other words, BRST quantization is equivalent to anti-BRST quantization for Yang-Mills theories \cite{Hull}. Really, this equivalence is clear in the case of Abelian gauge theories in which the BRST-anti-BRST relations are given as;
 \begin{eqnarray} \label {2-5-2}
\begin{array}{*{20}{c}}
   \delta A=-\emph{\emph{d}}C~~~~~~,~~~~~~~\widetilde{\delta}A=-\emph{\emph{d}}\widetilde{C},  \\
   \delta C=0~~~~~~~~~~~,~~~~~~~\widetilde{\delta} C=0,~~~~~  \\
   \delta \widetilde{C}=0~~~~~~~~~~~,~~~~~~~\widetilde{\delta} \widetilde{C}=0.~~~~~ \\
\end{array}
\end{eqnarray}
 
 \noindent These relations give rise to a symmetry between $(\delta,C)$ and $(\widetilde{\delta},\widetilde{C})$. This symmetry manifestly proves our claim above in the case of Abelian gauge theories.\\

 \noindent These consequences are also in complete agreement with our model for anti-BRST structure. To see this with more details recall that the anti-BRST derivative evaluates the variation of quantum fields with respect to infinitesimal transformations of gauge fixing function $f$, e.g. $f \to f'$. As mentioned above any transformation of $f$ leads to an infinitesimal gauge transformation, say $\alpha$, to shift the roots of $f$ to those of $f'$. On the other hand, any gauge transformation $\alpha$ applied to gauge fields can induce an infinitesimal transformation to $f$ conversely. This bijective relation provides an equivalence between gauge and gauge fixing transformations in Yang-Mills gauge theories. Now, it is not hard to see that this consequence is intimately correlated with isomorphism (\ref{2-7}).

%%%%%%%%%%%%%%%%%%%%%%%%%%%%%%%%%%%%%%%%

\par
\section{Summary and Conclusion}
\setcounter{equation}{0}
\par In this article we proposed a geometric model for anti-BRST structure of quanized Yang-Mills gauge theories based on the symmetry of classical gauge theories with respect to different gauge fixing methods. Here we showed that gauge fixing invariance is with enough evidences the classical counterpart of anti-BRST symmetry. Based on this model we also discussed the necessity of anti-BRST symmetry for any quantized BRST symmetric gauge theory.\\

\par In particular, once a classical gauge theory is given, its  independence of gauge fixing methods is naturally considered due to its physical consistency. Therefore, as it was shown above appearing anti-BRST symmetry in its quantized version must be as natural as its independence of gauge fixing mechanisms at classical level. In other words, finding anti-BRST symmetry in quantized version of a classical gauge theory is as relevant as appearing the BRST invariance in quantum Lagrangian. Consequently, all these results could be summarized in the following diagram;

\begin{eqnarray}  \label {2-0}
\begin{array} {*{20}{c}}
    {\underline{Classical~Symmetry}} & {~~~~~~Quantization~~~~~~} & {\underline{Quantum~Symmetry}} \\
   {\emph{\emph{Gauge~Invariance}}} & \to  & {\emph{\emph{BRST~Invariance}}}  \\
   { \updownarrow} & { } & { \updownarrow}  \\
   {\emph{\emph{Gauge~Fixing~Invariance}}} &  \to  & {\emph{\emph{Anti-BRST~Invariance}}}  \\
\end{array}
\end{eqnarray}
\\
\par Finally for completeness of our discussion we prefer to give some remarks. Firstly, it should be emphasized that although our model is arranged for strict gauge fixings or equivalently vanishing Nakanishi-Lautrup field BRST-anti-BRST transformations, it can be seen that the general form of BRST-anti-BRST structures for smooth gauge fixing mechanisms are exactly given with the same geometric background we discussed here \cite{Varshovi-next}. For the next remark, it must be noted that our model can be similarly promote to more general formalisms of gauge symmetries such as the Batalin-Vilkoviski approach \cite{Batalin-Vilkoviski} to provide more sophisticated geometric backgrounds for algebraic quantization methods. This will lead to anti-BRST structure in only on-shell closed and irreducible gauge theories. In other words, our geometric formulation of anti-BRST structure of quantized Yang-Mills theories, seems to have the ability to be accordingly generalized to other versions of gauge theories such as $p$-form gauge theories and the (super) gravity with the same background of gauge fixing symmetries in their own classical settings. Consequently we specify that it is not hard to see due to our model BRST and anti-BRST structures in Yang-Mills theories can be respectively considered as the first and the second entities of a hierarchical setting, the fact of which leads to a chain of generalized BRST formalisms in quantization of such gauge theories \cite{Varshovi-next2}.
\\

%%%%%%%%%%%%%%%%%%%%%%%%%%%%%%%%%%%%%%%%

%  with;
%  \begin{itemize}
%  \item $\emph{\emph{C}}^0(\mathbb{R}^m):=\{0\}$,
%  \item For $n=1$; $\emph{\emph{C}}^1(\mathbb{R}^m):=\{f\in C^\infty(\mathbb{R}^m)|f(0)=0\}$,
%  \item For $n=2$; $\emph{\emph{C}}^2(\mathbb{R}^m):=\{f\in C^\infty(\mathbb{R}^m\times \mathbb{R}^m)|f(p,0)=f(p,p)=0; p\in \mathbb{R}^m\}$,
%  \item For $n\geq3$; $\emph{\emph{C}}^n(\mathbb{R}^m)\subseteq C^\infty(\underbrace{\mathbb{R}^m\times ...\times\mathbb{R}^m}_{n- \emph{\emph{fold}}})$ consists of smooth functions $f$ with properties of $f(p_1,...,p_{n-1},0)=f(p_1,...,p_k,p,p,p_{k+1},...,p_{n-2})=0$, $k\leq {n-2}$, for any $p,p_1,...,p_{n-1}\in \mathbb{R}^m$,\\
%\end{itemize}
%  \noindent and for the linear maps

%%%%%%%%%%%%%%%%%%%%%%%%%%%%%%%%%%%%%%%%%%

%%%%%%%%%%%%%%%%%%%%%%%%%%%%%%%%%%%%%%%

%%%%%%%%%%%%%%%%%%%%%%%%%%%%%%%%%%%%%%%%%%%%%%%%%%%%%%%%%%%
%%%  \section{Harmonic Forms and Groenewold-Moyal Star Products}
%%%%%%%%%%%%%%%%%%%%%%%%%%%%%%%%%%%%%%%%%%%%%%%%%%%%%%%%%%%
%%     \setcounter{equation}{0}

%%%%%%%%%%%%%%%%%%%%%%%%%%%%%%%%%%%%%%%

%%%%%%%%%%%%%%%%%%%%%%%%%%%%%%%%%%%%%%%%%

%%%%%%%%%%%%%%%%%%%%%%%%%%%%%%%%%%%%%%%%%%%

 %%%%%%%%%%%%%%%%%%%%%%%%%%%%%%%%%%%%%%%%%%%

%%%%%%%%%%%%%%%%%%%%%%%%%%%%%%%%%%%%%%%%%%%

% \section{Conclusions}
%%%%%%%%%%%%%%%%%%%%%%%%%%%%%%%%%%%%%
% \par In this article 

%%%%%%%%%%%%%%%%%%%%%%%%%%%%%%%%%%%%%%%%%%%%%%%%
\section{Acknowledgments}
%%%%%%%%%%%%%%%%%%%%%%%%%%%%%%%%%%%%%%%%%%%%%%%%
\par The author says his special gratitude to S. Ziaee who was the main reason for appearing this article. Moreover, the author wishes to dedicate this article to all those who are trying for the holly wish of a global peace for all people and all nations all around the world including all those who belong to different genders, different races, different colors, different religions, different ideologies, different thoughts, and different philosophies with deep regards to humanity, since all of what we try to add to the human knowledge is meaningless without our common understanding of this humanity and an omnipresent peace. %Moreover, the author wishes to dedicate this article to all those who are trying for the holly wish of a global peace and prosperity for all people and all nations all around the world.

%The author should say his special gratitude to ... for use-full comments and fruit-full discussions. Moreover, my deepest thanks go to A. Shafiei Deh Abad for his kind considerations and his motivating ideas. Also I should confess that this article owes most of its appearance to S. Ziaee, whom my deepest regards goes to for many things.

%%%%%%%%%%%%%%%%%%%%%%%%%%%%%%%%%%%%%%%%%%%%%%%%%%%%%%%%%%%%%%%%%%
%\section{Appendices}
%%%%%%%%%%%%%%%%%%%%%%%%%%%%%%%%%%%%%%%%%%%%%%%%%%%%%%%%%%%%%%%%%%%
%\begin{appendix}\setcounter{equation}{0}\noindent
%\section{}\

%%%%%%%%%%%%%%%%%%%%%%%%%%%%%%%%%%%%%%%%%

\end{document}